# Autonomous Probe Microscopy with Robust Bag-of-Features Multi-Objective Bayesian Optimization: Pareto-Front Mapping of Nanoscale Structure–Property Trade-Offs

Kamyar Barakati[1,*], Haochen Zhu[1], C Charlotte Buchanan[1], Dustin A Gilbert[1], Philip Rack[1,†], and Sergei V. Kalinin[1,2,‡]

[1] Department of Materials Science and Engineering, University of Tennessee, Knoxville, TN 37996

[2] Pacific Northwest National Laboratory, Richland, WA 99354

**Abstract:**

Combinatorial materials libraries are an efficient route to generate large families of candidate compositions, but their impact is often limited by the speed and depth of characterization and by the difficulty of extracting actionable structure–property relations from complex characterization data. Here, we develop an autonomous scanning probe microscopy (SPM) framework that integrates automated atomic force and magnetic force microscopy (AFM/MFM) to rapidly explore magnetic and structural properties across combinatorial spread libraries. To enable automated exploration of systems without a clear optimization target, we introduce a combination of a static physics-informed bag-of-features (BoF) representation of measured surface morphology and magnetic structure with multi-objective Bayesian optimization (MOBO) to discover the relative significance and robustness of features. The resulting closed-loop workflow selectively samples the compositional gradient and reconstructs feature landscapes consistent with dense grid "ground truth" measurements. The resulting Pareto structure reveals where multiple nanoscale objectives are simultaneously optimized, where trade-offs between roughness, coherence, and magnetic contrast are unavoidable, and how families of compositions cluster into distinct functional regimes, thereby turning multi-feature imaging data into interpretable maps of competing structure–property trends. While demonstrated for $Au - Co - Ni$ and AFM/MFM, the approach is general and can be extended to other combinatorial systems, imaging modalities, and feature sets, illustrating how feature-based MOBO and autonomous SPM can transform microscopy images from static data products into active feedback for real-time, multi-objective materials discovery.

[*] kbarakat@vols.utk.edu
[†] prack@utk.edu
[‡] sergei2@utk.edu

**I. Introduction**

Combinatorial materials libraries were introduced as a route to accelerate materials discovery and optimization by transitioning from single composition at a time to continuous spreads or dense arrays.[1] Early studies demonstrated that thin-film co-deposition, gradient growth, and related methods could generate hundreds to thousands of distinct compositions on a single substrate, enabling systematic mapping of structure–property relationships in ferroelectric oxides[2], phosphors[3], catalysts[4], and functional alloys[5]. However, it was immediately realized that while combinatorial platforms provide an efficient way to generate candidate materials sampling quazi-binary or quasi-ternary cross-sections of phase diagrams, their impact is ultimately determined by how effectively those libraries can be read out and interpreted.[6]

The central bottleneck has therefore shifted from synthesis to characterization and analysis.[7] While structural and optical probes such as diffraction, reflectometry, and hyperspectral imaging have been adapted for wafer-scale or array-scale mapping, many key functional properties, including transport, catalytic activity[8], ferroic responses[9], and local reactivity, remain slow and experimentally demanding. When high-throughput measurements are available, they frequently yield complex data objects (images, spectra, time series) that require substantial expert intervention to process and interpret.[10] As a result, combinatorial libraries are often rich in samples but sparse in actionable information: large regions of composition–process space can be generated, but only a fraction is interrogated with sufficient depth, and the resulting data rarely feedback directly into subsequent synthesis and measurement decisions.[11, 12]

A third wave of combinatorial discovery is now emerging, built around the integration of machine learning for complex materials[13-17], Bayesian optimization[18-20], and robotic automation[21-23] into adaptive experimental loops. In this setting, libraries are no longer treated as static arrays to be screened exhaustively, but as landscapes that can be explored selectively, with models updated as measurements are acquired and used to propose the next experiments. Most current implementations, however, rely on relatively simple scalar readouts and techniques that allow rapid measurements, enabling implementation of Bayesian Optimization based active experimentation.[22, 24] Among these, scanning probe microscopy (SPM) which can access local structure, functionality, and heterogeneity that often control macroscopic behavior offer a promising paradigm for the realization of closed-loop combinatorial workflows.[25, 26] However, the challenge in this case is often the transition from multidimensional data sets such as images and spectra to the reward functions that can be used as an exploratory target for active learning.

In this work, we address this gap by developing an autonomous SPM workflow for combinatorial libraries and coupling it to a "bag-of-features" (BoF) multi-objective Bayesian optimization (MOBO) scheme.[27] Using the ternary Au-Co-Ni composition-spread as a model system, we integrate automated atomic force microscopy (AFM) with a control layer that can position, image, and analyze nanoscale regions across the library without human intervention. AFM images are transformed into feature representations that capture relevant aspects of morphology and local response, and these feature vectors serve as multi-dimensional objectives within a MOBO framework. The optimizer selects new measurement locations in composition space based on predicted performance and uncertainty in this feature space, enabling efficient, adaptive exploration of the library. Importantly, the discovered Pareto frontiers illustrate the relationships in original BoF targets, closing the discovery loop. In this way, we show how combinatorial libraries, autonomous SPM, and feature-based Bayesian optimization can be combined into a unified workflow in which nanoscale readouts function as active feedback, continuously informing real-time decisions in materials discovery.

**I. Materials and ground truth measurements**

As a testbed for autonomous scanning probe workflows on combinatorial libraries, we seek a system that combines a well-established bulk phase diagram, clear functional contrast, and sensitivity to nanoscale microstructure.[28, 29] Magnetic alloys meet these requirements particularly well: their macroscopic properties depend strongly on composition, emergent phases, morphology, and domain structure, while their local behavior can be interrogated directly by SPM.[30-32] Within this class, ternary systems offer an additional degree of freedom, enabling continuous tuning of exchange interactions and anisotropy, as well as the possibility of emergent phases not present in the binaries.[33-35] These features make magnetic ternaries a natural choice for demonstrating how adaptive, feature-based exploration can uncover complex structure–property relationships across a composition spread.

Here, we focus on the $Au - Co - Ni$ ternary as a model system. $Co$ and $Ni$ are both ferromagnetic 3d transition metals[36], but they exhibit distinct magnetic characteristics. $Co$ is associated with comparatively high magnetic anisotropy and robust ferromagnetism[37], while $Ni$ behaves as a softer ferromagnet with different domain structures and magnetoelastic responses[38]. $Au$, in contrast, is a noble, nominally non-magnetic metal that is expected to act primarily as a diluent and structural modifier, influencing grain size, segregation tendencies, and local electronic environments. Additionally, Au is plasmonically active, thus this ternary system can be tuned for intriguing bi-functional magneto-optical properties.[39-42] The bulk $Au - Co - Ni$ phase diagram and many of its binary subsystems have been extensively mapped, providing a reference for equilibrium phases, solubility ranges, and magnetic trends as a function of composition.[41] While the $Co - Ni$ binary is nearly isomorphous and thus contains a large mutual solubility range, the $Au - Ni$ and $Au - Co$ are insoluble due to the large size difference. This existing knowledge base allows us to distinguish genuinely new behaviors from those that can be anticipated from bulk thermodynamics.

Combinatorial thin-film growth, however, can access regimes that differ substantially from bulk equilibrium. In particular, the rapid quenching of the condensed atoms leads to meta stable solid solutions with microstructures and magnetic properties that deviate from bulk expectations.[43] In the $Au - Co - Ni$ library, we therefore anticipate the formation of composition-dependent morphologies with varying grain sizes, roughness, and phase distributions, as well as spatially heterogeneous magnetic responses that are sensitive to local composition and microstructure. Automated AFM and magnetic-force-based measurements are well suited to capture this behavior: they provide simultaneous access to surface topography and local magnetic contrast across the spread. Within our autonomous workflow, these nanoscale readouts become the primary observables used to characterize the library and to guide the multi-objective Bayesian optimization of measurement locations in composition space.

To establish a quantitative baseline for subsequent optimization studies, a grid-based AFM/MFM characterization campaign was conducted across the 10-cm-diameter $Au - Co - Ni$ gradient film. A total of 33 uniformly distributed positions were measured across the substrate, each using a 256 × 256-pixel scan with 1 $\mu m$ lateral resolution, corresponding to approximately 8 minutes per acquisition. Prior to the AFM/MFM, the sample was mapped using both X-ray diffraction (XRD), scanning electron microscopy, and energy dispersive X-ray spectroscopy (EDS) to map the phases present, surface topology and composition, respectively. **Figure 1** illustrates the spatially resolved composition and magnetic-property landscape as a function of position across the 100 mm diameter substrate as well as the resultant thickness gradient. Magnetization per atom was estimated using a simple dilution model, with the experimental compositions, and the bulk moments of 1.7 $\mu_B$/atom and 0.6 $\mu_B$/atom for Co and Ni, respectively. This is likely a slight over-estimation since Au suppresses the magnetization beyond dilution effects through broadening of the d-bands and spin-orbit effects. This wafer-scale mapping establishes the experimental parameter space from which composition–property relationships are extracted and subsequently explored using grid-based and optimization-guided sampling strategies. Surface topography measured using scanning electron microscopy (SEM) and corresponding X-ray diffractograms acquired from five representative locations are presented in **SI Section SIII (Figure S5)**. The diffraction data demonstrate that the as-deposited films form metastable solid-solution phases, while the SEM images reveal a nanogranular film morphology, consistent with rapid solidification during room-temperature deposition. EDS mapping further confirms the absence of detectable phase separation across the library (see **SI Section SIII, Figures S4 and S6, and Table S1** for data summarized from **13** locations).

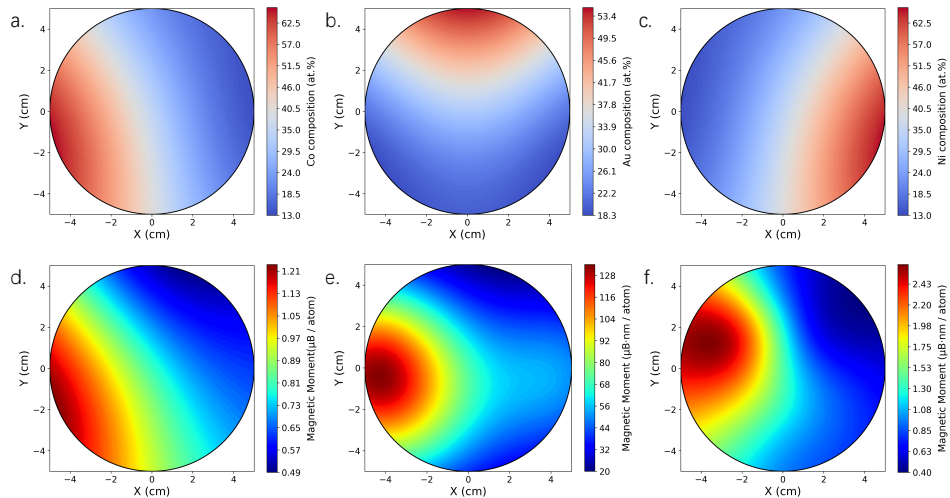

*Figure 1*: Spatially resolved maps of composition and magnetization across the Co–Ni–Au combinatorial wafer. Panels (a–c) show the interpolated atomic composition distributions of (a) Co, (b) Au, and (c) Ni, expressed in atomic percent and mapped as a function of wafer position (X, Y). Panels (d–f) display the corresponding magnetic properties derived from the same compositional library: (d) calculated magnetization per atom ($\mu_B$ atom$^{-1}$), and (e) magnetization normalized by film thickness ($\mu_B$·nm atom$^{-1}$), and (f) experimentally measured magnetic moment from magnetometry. All maps are interpolated from discrete measurement points and masked to the circular wafer geometry.

This procedure produced a spatially resolved library of nanoscale morphological and magnetic information that can be correlated to the composition of the $Au - Co - Ni$ metastable solid solution.

Realizing active learning strategies requires scalar or several scalar descriptors that can be used as optimization targets. For measurements such as nanoindentation, these can be natively chosen as measured scalar parameter (e.g. hardness).[44] However, for multidimensional spectroscopic and imaging studies, the choice of optimization target (scalarizer) can be non-trivial. For example, for X-ray diffraction (XRD) based active learning, exploration strategies can be formulated based on phase lattice parameters, phase compositions, or disorder estimates, leading to fundamentally different exploration results. For spectroscopic measurements in electron and scanning probe microscopy,[45-48] construction of scalarizer is the primary factor determining the automated experiment progression and physical relevance of derived results. Since the prior knowledge of the scalarizer can be uncertain, the human in the loop strategies and novelty exploration approaches have been introduced to minimize the possible biases.[49-53] However, for imaging data the construction of scalarizer functions that convert images to the optimization targets have not yet been explored.

Here, we propose an approach based on the bag-of-features, when multiple physically relevant robust scalarizers are constructed and used for multi-objective optimization, dynamically learning the relationships between them. For the imaging data, robustness refers to the scalarizer stability with respect to the choice of the sampling grid and translations within the image plane (and ideally imaging conditions). For physical relevance, the scalarizers should serve as a proxy to the discovery objective. Here, from each AFM and MFM measurement, a set of physically interpretable descriptors was extracted to quantify both surface morphology and magnetic response. The complete set of metrics and their physical interpretation is presented in **Table 1**, with the corresponding computational procedures detailed in **SI Section SII**. These features constitute the ground-truth response surfaces shown in **Figure 2**, which together capture the continuous yet heterogeneous variation in particle size, structural correlation length, magnetic domain size, and domain magnitude across the composition-gradient substrate.

*Table 1*: *Summary of extracted AFM and MFM features*

| Feature Name | Channel | Physical Meaning |
| --- | --- | --- |

| Mean Particle Diameter | AFM Height | Measures average lateral size of surface grains or particles; reflects nucleation and growth behavior. |
| Correlation Length | AFM Height | Quantifies the characteristic lateral spacing of surface features; related to surface ordering and texture. |
| Magnetic Domain Size | MFM Phase | Represents the lateral size of magnetically coherent regions; linked to domain structure and magnetic anisotropy. |
| Magnetic Domain Magnitude (Intensity) | MFM Phase | Reflects the strength of local magnetic contrast; sensitive to magnetic moment variations and domain orientation, particularly along the film normal direction. |

The spatial variations observed across the feature maps in **Figure 2**, arise directly from the underlying physics of the $Au - Co - Ni$ composition-gradient thin film. The local alloy composition changes smoothly across the wafer, and this variation governs both the surface morphology and the magnetic microstructure. The AFM-derived descriptors show gradual changes because differences in local Au, Co, and Ni content influence surface diffusion, nucleation density, and island coarsening during growth, leading to continuous shifts in particle size and the lateral spacing of surface features. In contrast, the MFM-based descriptors exhibit sharper spatial transitions, reflecting the strong dependence of magnetic domain structure on the relative proportions of Co and Ni, the primary ferromagnetic constituents in the alloy. Changes in their local concentration modify the effective magnetic moment, exchange interactions, and the energetics of domain formation, producing distinct variations in both domain size and magnetic signal strength across the substrate. Taken together, these experimentally measured landscapes constitute a high-fidelity benchmark against which automated exploration and model-based optimization strategies can be rigorously validated.

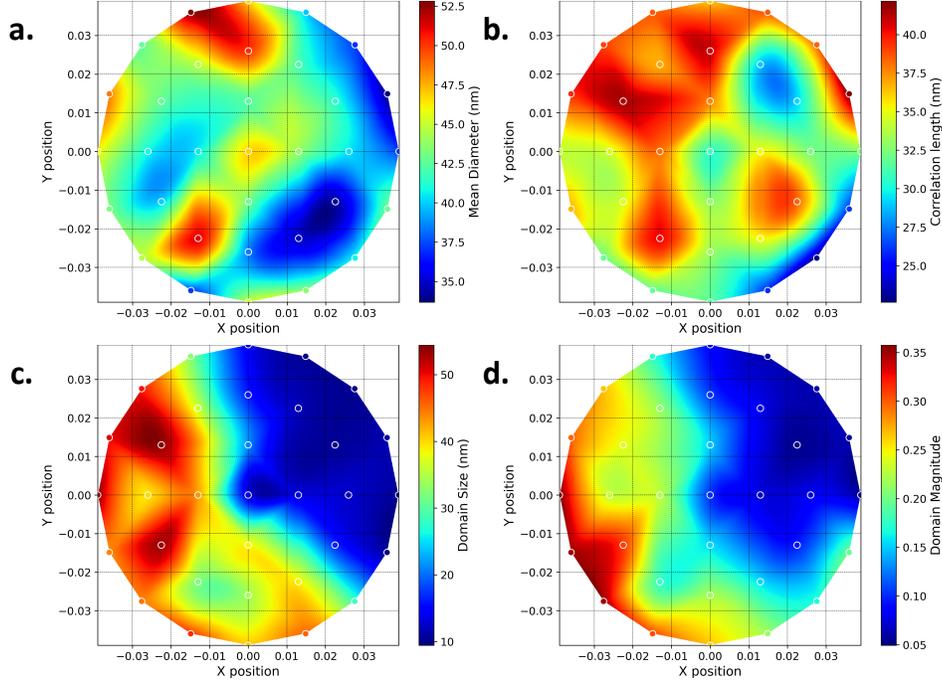

*Figure 2*: *Ground-truth objective landscapes obtained via grid search across the sample surface, where: (a) mean diameter, (b) correlation length, (c) domain size, and (d) domain magnitude.*

## II. Framework for Intelligent Probe Microscopy

Exploration of combinatorial libraries in which multiple aspects of material behavior vary simultaneously across composition–process space remains relatively underdeveloped. When a single target property is known in advance, for example, a specific magnetic figure of merit, a mechanical hardness, or an electrical conductivity, the problem can be cast naturally in the framework of classical Bayesian optimization. In this setting, a scalar objective function is defined over the experimental parameter space, and a probabilistic surrogate model is used to approximate this function from a limited number of measurements. The surrogate (typically a Gaussian process, structured Gaussian Process, or any model that yields prediction and uncertainty) plays a central role in this workflow: given measured data, it interpolates the objective across the parameter space and provides an associated uncertainty estimate. An acquisition function constructed from the surrogate mean and uncertainty then guides the active learning process, selecting new experimental conditions by balancing exploitation of promising regions with exploration of uncertain ones.

For combinatorial libraries where multiple functionalities are present and the "right" objective is not known a priori, this scalar-optimization picture is no longer adequate. In such cases, the goal is not simply to maximize a single property, but to discover and map diverse modes of behavior including morphological, magnetic, electronic, that may emerge in different regions of the library.

To address this challenge, we introduce an approach based on a "bag-of-features" representation coupled to multi-objective Bayesian optimization (MOBO). Here, the bag-of-features is defined as a collection of descriptors derived from combined AFM and MFM images, designed to jointly encode topographic and magnetic characteristics at the nanoscale. These descriptors are chosen to be robust: they should be computable reliably from imaging data, insensitive to minor variations in sampling or resolution, and, as far as possible, invariant to trivial transformations such as translations. Descriptors constructed in such way are not independent and can have strong correlations across the exploration space. Similarly, these descriptors are intended to serve as proxies for the functionality of interest (exploration objective) that is not directly observed during combinatorial exploration; however, this relationship is unknown before the experiment. A comprehensive description of the model architecture, feature construction, and encoding strategy is provided in **SI Section SI, and SII.**

Once the bag-of-features representation is defined, the exploration problem is reformulated as a multi-objective optimization in feature space. Each measurement location in the combinatorial library yields a vector of feature values, and MOBO is used to construct surrogate models and acquisition strategies that operate on this vector-valued response. The optimizer then selects new points in composition–process space that are expected to improve coverage and performance across the feature objectives, enabling systematic discovery of distinct morphological and magnetic regimes. The discovered Pareto fronts then establish the relationships between the descriptors in an active learning setting. In principle, this framework can be extended to include additional criteria such as novelty, diversity, or user-defined reward functions. In the present work, however, we focus on the core MOBO formulation, demonstrating that a bag-of-features representation is sufficient to enable adaptive, feature-level exploration of multifunctional combinatorial libraries.

## II.A. Bag-of-feature construction

A central challenge in autonomous exploration of combinatorial libraries is that we do not know a priori what the "right" descriptors are for a given material system or functionality. Any fixed choice of features risks over-committing to a particular hypothesis about what aspects of the images matter most. Instead, we treat descriptors in a probabilistic manner: rather than relying on a single scalar metric, we construct a bag-of-features (BoF) representation in which each measurement is mapped to a vector of candidate descriptors, and the subsequent Bayesian optimization operates over this feature space and its associated uncertainties. In this view, the surrogate models and acquisition strategies statistically determine which descriptors, individually or in combination, are most informative, eliminating the need for manual feature selection.

Within this framework, the design of the BoF must meet several key requirements stemming from the nature of phenomena of interest and type of data. First, the features must be relevant to the underlying materials functionalities of interest. For AFM/MFM data, this means that the descriptors should capture salient aspects of morphology (roughness, grain size distributions, texture measures) and magnetic behavior (domain patterns, contrast statistics, correlation lengths) that can plausibly connect to macroscopic performance. Second, the BoF must be robust with respect to sampling conditions. In an adaptive autonomous workflow, we do not know in advance the "correct" resolution, image size, or sampling strategy for each region of the library; different areas may be imaged at different zoom levels or pixel densities. Descriptors should therefore be constructed so that modest changes in resolution or field of view do not qualitatively alter their values, allowing meaningful comparison across the library.

A third requirement is robustness to translational invariance. Microscopy images typically lack an intrinsic point of origin, and the features can appear anywhere within the field of view without changing the underlying microstructure. Descriptors that depend sensitively on absolute position within the frame are thus undesirable. Instead, the BoF should emphasize statistics and patterns that are invariant (or only weakly sensitive) to translations, so that the same physical structure yields similar descriptors regardless of where it falls in the scan window. Together, probabilistic treatment of descriptors and these constraints, functional relevance, sampling robustness, and translational invariance, ensure that the BoF provides a stable, physically meaningful basis for multi-objective Bayesian optimization across the combinatorial library.

The combined requirements of functional relevance, sampling robustness, and translational invariance ensure that the BoF representation provides a stable and physically grounded basis for multi-objective Bayesian optimization across the combinatorial library. Guided by these criteria, the selected descriptors were chosen to capture the dominant structural and magnetic signatures present in the AFM and MFM data. The resulting feature set, comprising Fourier-derived magnetic domain size, autocorrelation-based correlation length, Voronoi-extracted particle diameter and area, and a contrast-based magnetic domain magnitude, encodes complementary aspects of morphology and magnetic response. These descriptors, summarized in **Table 1** and detailed in **SI Section SII**, provide a compact and physically interpretable representation of the underlying behavior of the gradient film.

**II.B. Multi-objective optimization and Pareto hypothesis testing**

With the bag-of-features representation in place, we cast the problem of discovery over the combinatorial library as a multi-objective Bayesian optimization (MOBO) task. For vector-valued responses, there are multiple possible exploration strategies: one can, for example, prioritize novelty by seeking measurements that produce feature vectors far from previously observed ones, or adopt information-theoretic criteria that emphasize uncertainty reduction in specific regions of feature space. Several such strategies have been proposed in the context of multi-objective and goal-free discovery, where the emphasis is on identifying diverse and unexpected behaviors rather than optimizing a single predefined metric.

Here, we adopt a more classical multi-objective optimization strategy in which each feature (or feature group) is treated as an objective in a joint MOBO framework. Given multiple objectives, the goal is not to identify a single "best" point but to approximate the Pareto frontier: the set of points in parameter space for which none of the objectives can be improved without degrading at least one other. Multi-objective Bayesian optimization uses surrogate models to approximate each objective and its uncertainty, and acquisition functions defined in terms of dominance, hypervolume improvement, or related measures to propose new experimental conditions that move the current approximation toward the true Pareto front.

Combined with the bag-of-features representation, the Pareto frontier provides a natural lens on how different nanoscale descriptors co-vary across the library. If the Pareto set collapses to a single small region in composition–process space, this indicates that all feature objectives are simultaneously optimized there, suggesting a narrow optimum. By contrast, an extended Pareto front implies trade-offs among features. For example, improvements in magnetic contrast at the expense of increased roughness or vice versa will highlights families of compositions that realize different balances between functionalities. As such, MOBO on a feature-based representation is well suited for exploring combinatorial libraries in which the goal is to map and exploit relationships among multiple functionalities, rather than to maximize a single predefined property.

In this framework, the MOBO is employed to guide autonomous exploration when multiple quantitative objectives govern the experimental landscape. Properties such as magnetic domain morphology, compositional uniformity, and magnetic signal magnitude are modeled as distinct response surfaces using GP surrogates, with each GP capturing both the predicted value and the uncertainty associated with each objective. Their joint posterior defines a probabilistic representation of the trade-off surface, or Pareto front, across the compositional space. To select new measurements, several acquisition formulations can be considered. Classical single-objective criteria such as Expected Improvement (EI) and Upper Confidence Bound (UCB) are extended to the multi-objective setting to quantify either the expected gain over current optima (EI-based approaches) or a principled uncertainty–exploration balance (UCB-based approaches). In the present work, the acquisition is performed using q-Expected Hypervolume Improvement (qEHVI), which evaluates the expected increase in the dominated hypervolume of the Pareto set from a batch of proposed experiments. This criterion directs sampling toward regions most likely to expand the Pareto-optimal front and reveal informative trade-offs among competing objectives. By prioritizing measurements that maximize the expected hypervolume gain, qEHVI provides an effective strategy for autonomous navigation of the multi-objective landscape defined by the BoF descriptors. Full mathematical details of the GP construction, kernel parameterization, and the qEHVI acquisition function are provided in **SI Section SI**.

**Implementation and Performance of MOBO**

The automated MOBO exploration was implemented in a Python environment using the BoTorch library, which provides Gaussian Process (GP) surrogate models and multi-objective acquisition functions suitable for closed-loop experimentation. The optimization was performed over the compositionally graded $[Au]\_x [Co]\_y [Ni]\_(1 - x - y)$ thin-film sample, previously characterized in **Section I**, using the objective functions defined in **Section II**. The algorithm was initialized with an initial seeding set of $N = 10$ randomly selected measurement locations uniformly distributed across the sample surface to establish a prior GP model for each objective. Subsequent acquisition points were selected sequentially according to the qEHVI criterion, iteratively refining the GP posteriors over a total of $N = 25$ optimization steps.

In **Figure. 3a**, the reconstructed mean diameter map reveals a compositional trend consistent with the ground-truth morphology presented in **Figure 2a**, though shifted toward the Au- and Co-rich regions of the film. This displacement indicates that surface coarsening is energetically favored in both compositional extremes, arising from distinct underlying mechanisms. In $Au-$enriched zones, the enhanced adatom mobility promote lateral grain growth and coalescence, while in $Co-$rich regions, phase segregation and magnetic clustering likely drive grain boundary rearrangement, resulting in similarly enlarged surface features. The optimization trajectory rapidly converged toward these coarser domains, underscoring the model's ability to identify and exploit multiple local optima within a complex morphological landscape.

The corresponding correlation-length distribution in **Figure. 3b** provides a complementary measure of spatial coherence, revealing that regions with larger particle areas exhibit proportionally longer correlation lengths. The highest $\xi$ values occur along the $Co-Au$ compositional boundary, where enhanced surface diffusion produces smooth, laterally continuous textures, whereas $Ni-$rich compositions display shorter correlation lengths, reflecting dense nucleation, fine-scale roughness, and limited lateral order; the formal definition of $\xi$ is given in **SI Section SII**. This spatial modulation in $\xi$ encapsulates the underlying thermodynamic and kinetic asymmetry of the alloy system, linking atomic mobility and surface energy to emergent morphological organization. Collectively, these reconstructions demonstrate that the Bayesian optimization framework captures both the amplitude and spatial correlation of topographic features, effectively mapping the coupled structure–composition relationships that govern surface evolution in multicomponent materials.

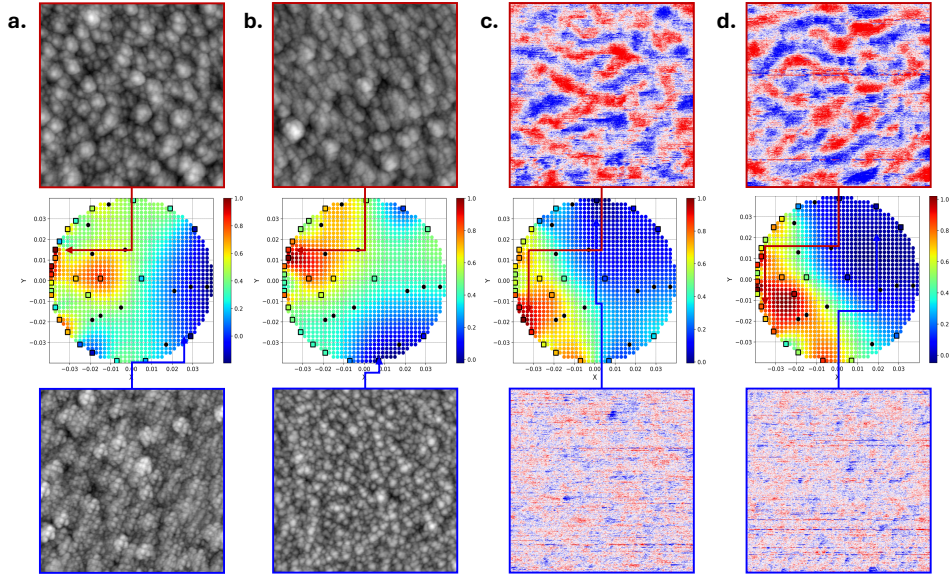

*Figure 3*: MOBO-derived maps after N optimization steps, showing the spatial distributions for each objective over a 1 μm × 1 μm scanned area: (a) mean diameter, (b) correlation length, (c) magnetic domain size, and (d) magnetic domain magnitude. Highlighted magnified regions marking the locations of maximum (top row) and minimum (bottom row) solutions (positions on the sample) for each objective upon completion of the optimization.

The reconstructed magnetic domain size map in in **Figure. 3c** reflects the dominant spatial periodicity of the magnetic phase patterns, as defined in **Eq. (1) in SI Section SII.** The characteristic length scale $\mathcal{L}$ corresponds to the inverse of the principal spatial frequency $f^*$ obtained from the azimuthally averaged Fourier power spectrum of the MFM phase images. Regions with larger $\mathcal{L}$ exhibit coarse and widely spaced domains, while smaller $\mathcal{L}$ values indicate fine, densely packed magnetic textures. In the present compositional landscape, the largest domain sizes appear near the $Co-rich$ boundary, reflecting the strong exchange stiffness which is known to exist within Co, relative to Ni, and the low pinning density from the low Au content. Moving towards the $Au-rich$ edge of the wafer, the non-magnetic element interrupts the long-range exchange interaction leading to much smaller domains. The Au also introduces a large density of pinning sites which constrain domain propagation by arresting the movement of the domain walls, resulting in many smaller domains, as observed in the MFM images. Notably, these domains are much smaller than the thickness of the film, suggesting that the structures on the Au-edge of the sample likely do not extend through the film, as the labyrinth domains on the Co-edge likely do. Thus, the MFM reveals changes in the domain morphology reflective of the composition and atomic-scale compositional mixing (i.e. the exchange interaction) and the microscale structure (the pinning potential). It is also important to note that the domain size is determined by the balance of the exchange energy versus the magnetostatic energy, which scales with the out-of-plane magnetization, with higher moments causing smaller domains. In the current data, regions with the highest magnetization also have the largest domains, contrary to this conventional understanding, however, this simply indicates the suppressive effects on the exchange interaction are larger than the dilutive reduction of the magnetization. The corresponding magnetic domain magnitude (intensity) distribution shown in **Figure. 3d**, derived from **Eqs. (5–9)** in **SI Section SII**, quantifies the local magnetic contrast $R\_mag$ between upward and downward domains. Regions with higher $R\_mag$ correspond to stronger attraction/repulsion between the material and the MFM cantilever. Assuming the magnetic moments are oriented in the out-of-plane orientation, this signal is a direct proxy for the saturation magnetization; relaxing that assumption, low $R\_mag$ may indicate reduced magnetization due to dilution with Au or poor ordering of the Co and Ni – which results in band broadening – or orientation of the domains away from the out-of-plane direction, i.e. in-plane relaxation. The highest intensities occur along the $Co-rich$ region and decreases towards the Au region.

In addition, the spatial trends observed in both the reconstructed domain size and magnetic domain magnitude maps are fully consistent with the theoretically predicted magnetic moment distributions shown in **Figure 1**. The $\mu_B$/atom map reflects intrinsic, composition-dependent magnetization, while the $\mu_B \cdot$nm/atom map accounts for the effective magnetic volume contribution of the film. Regions identified by MOBO as exhibiting larger magnetic domain sizes and stronger magnetic contrast correspond closely to areas of elevated effective magnetic moment, confirming that the optimized MFM objectives capture physically meaningful magnetic variations rather than stochastic imaging features.

This consistency is further supported by the compositional gradients resolved by EDS (**Figures S4** and **S6**; **Table S1**), which reveal smooth lateral changes in Au, Co, and Ni content across the sample. The absence of abrupt compositional discontinuities ensures that the observed magnetic-domain evolution arises from gradual changes in exchange interaction strength and magnetic dilution rather than from chemical phase separation.

Location-dependent XRD patterns (**Figures S5** and **S7**) exhibit broad diffraction maxima near $2\theta \approx 40°$ and $70°$, consistent with FCC (111) and (220)–type reflections typical of chemically disordered Au–Co–Ni alloys. Importantly, the peak positions and overall diffraction profiles vary smoothly with location and composition, with no appearance of new reflections or abrupt structural changes indicative of crystallographic phase transitions.

Accordingly, the observed spatial evolution of magnetic domains cannot be attributed to changes in crystal symmetry or the formation of distinct crystalline phases. Instead, the magnetic behavior is primarily governed by composition-driven variations in local exchange interactions, pinning site density, and saturation magnetization, arising from changes in chemical composition and nanoscale structural heterogeneity rather than from differences in grain orientation or long-range crystalline anisotropy. Jointly, these results establish a clear correspondence between MOBO-derived magnetic objectives and conventional microstructural, compositional, and magnetization measurements, demonstrating that multi-objective MFM provides a reliable, high-throughput proxy for mapping complex magnetic materials while remaining grounded in experimentally validated physical properties.

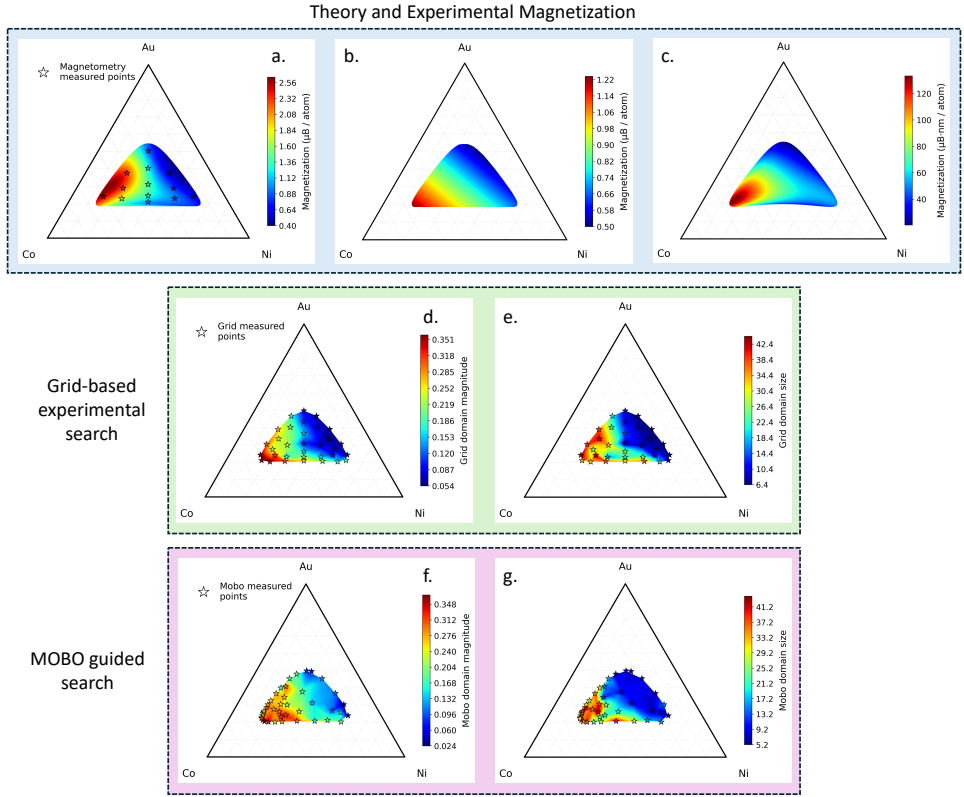

*Figure 4*: Exploration and optimization in compositional space. Ternary phase diagrams showing properties mapped in compositional space (Co–Ni–Au), where each point represents a unique alloy composition ( $[Co]_x [Ni]_y [Au]_{(1-x-y)}$ ). The triangular coordinates correspond to atomic fractions of Co, Ni, and Au, and the shaded region indicates the subset of the full ternary diagram that is experimentally accessible in the combinatorial library. **a)** Magnetization per atom measured by magnetometry at selected alloy compositions, normalized per atom ($\mu_B$ atom$^{-1}$). **b)** Theoretically predicted magnetization per atom as a continuous function of composition, evaluated using experimentally measured local alloy compositions. **c)** Effective magnetization per atom normalized by surface area, incorporating experimentally measured local composition and film thickness. **d)** Discrete grid-based experimental sampling of the predicted magnetization landscape, using domain magnitude as a proxy, with star markers indicating measured compositions and colors representing the corresponding magnetization values. **e)** Magnetic domain size mapped across compositional space using the same grid-based sampling strategy. **f)** Magnetization values obtained via multi-objective Bayesian optimization (MOBO), with star markers indicating compositions selected and measured during the optimization process. **g)** Domain size corresponding to the MOBO-selected compositions using the same MOBO strategy.

To place the magnetic behavior of the $Au - Co - Ni$ system in a broader compositional context, ternary phase diagrams are used to simultaneously visualize simulated expectations and experimentally measured responses. **Figure 4a** presents the experimentally measured magnetic moment per atom obtained from magnetometry at **13** discrete alloy compositions (**Table S1**), providing a direct experimental sampling of the magnetic response across the compositional space.

The simulated maps shown in **Figures 4b and 4c** establish an a priori description of the magnetic landscape, depicting the variation of the magnetic moment ($\mu_B$ atom$^{-1}$) and the magnetic moment–thickness product ($\mu_B$·nm atom$^{-1}$) across the accessible composition range. Note that the MFM probes near-surface magnetic behavior and is insensitive to film thickness; therefore, it is most appropriately compared to the magnetization per atom shown in **Figure 4c**. In contrast, magnetometry measures the total magnetic moment of the film and thus scales with the volume of magnetic material, making it representative of the magnetization–thickness product shown in **Figure 4b**. These simulations and actual magnetometry measurements provide physical intuition about how magnetism is expected to evolve with alloying and serve as a reference against which experimental exploration can be assessed. When the compositions sampled during MOBO are overlaid on these maps, it becomes evident that the optimization trajectory preferentially probes specific regions of compositional space rather than sampling uniformly, reflecting the structure of the underlying magnetic landscape. Figures 4d and 4e present experimentally measured magnetic domain magnitude and domain size obtained via uniform sampling, providing an independent experimental characterization of the magnetic response across composition. As expected, good agreement is observed between the magnetometry measurements shown in **Figure 4a** and the domain-based magnetic measurements in **Figure 4c**.

The magnetic domain magnitude and domain size measured at compositions selected during MOBO are shown in **Figures 4f** and **4g**, respectively, mapped directly within the compositional space. These results show that the optimization trajectory explores specific regions of compositional space rather than sampling uniformly, in contrast to the uniformly sampled measurements presented in **Figures 4d** and **4e**. This preferential exploration reflects the structure of the underlying magnetic landscape, with the MOBO algorithm adaptively guiding measurements toward compositions associated with enhanced magnetic response. The similar localization of MOBO-selected compositions observed for both magnetic domain magnitude and domain size indicates that the optimization captures correlated features of the magnetic behavior across the accessible compositional space. To further validate this observation, magnetometry measurements were conducted across the film library. The results in **Figures 4a** and **4g** show good agreement, supporting the use of MFM as a convenient proxy for assessing the magnetic behavior of the thin films. Although this grid-based approach ensures broad compositional coverage, the resulting measurements are necessarily sparse and provide only a coarse representation of the response surface, limiting the ability to resolve localized extrema or narrow regions of enhanced performance. This limitation becomes particularly apparent when compared to **Figures 4f** and **4g**, where MOBO-driven experiments are mapped in the same compositional coordinates. Here, the measured points cluster in regions associated with favorable magnetic responses, indicating that MOBO efficiently steers experiments toward high-performing compositions while avoiding less informative regions. Taken together, this comparison highlights how adaptive optimization strategies can more effectively navigate complex compositional spaces, revealing structure in the response landscape that is difficult to capture through uniform sampling alone and enabling a more efficient identification of optimal magnetic compositions.

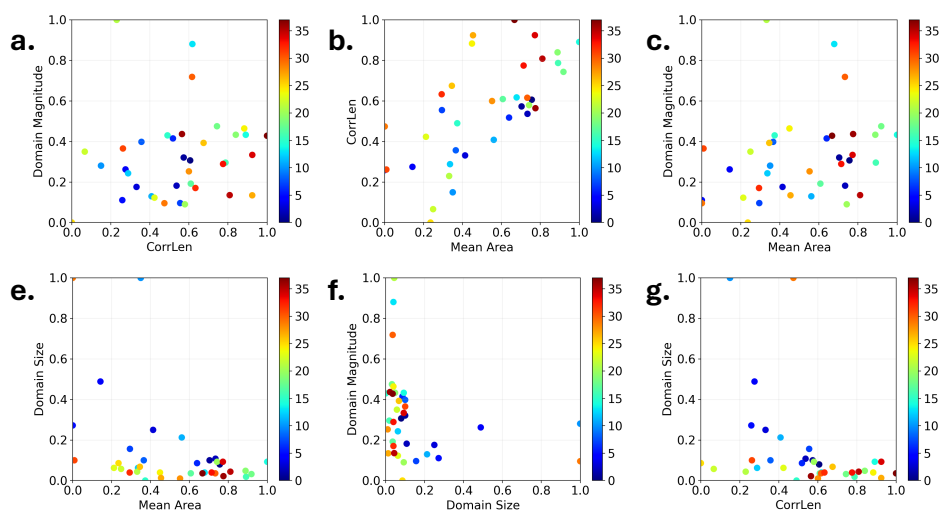

*Figure 5*: *Scatter plots illustrating Pareto-front solutions between different pairs of objectives obtained from multi-objective Bayesian optimization (MOBO). Each point represents a Pareto-optimal solution (scanned area of the sample), while the color bar indicates the MOBO iteration steps, capturing the progression of the optimization process. The plots highlight trade-offs among objectives such as domain magnitude, mean area, correlation length, and domain size.*

Building upon these spatial reconstructions, **Figure 5,** presents the Pareto-front distributions among the four objectives, *mean diameter*, *correlation length*, *magnetic domain size*, and *magnetic domain magnitude*, revealing how the optimization process explores and reconciles the trade-offs between morphological and magnetic properties. Each data point represents an experimentally evaluated configuration, color-coded by iteration to trace the evolution of the optimization trajectory.

Initially, the solutions are widely scattered across the parameter space, reflecting random exploration and limited correlation among objectives. With successive iterations, the distributions contract and align along well-defined Pareto fronts, indicating that the algorithm has learned the underlying interdependencies between structural and magnetic responses. The progression toward the upper-right frontier signifies a systematic improvement of both feature coherence and magnetic performance, where smoother, coarser morphologies coincide with larger, more magnetically distinct domains. This convergence behavior demonstrates that the multi-objective Bayesian framework not only identifies high-performing regions but also captures the intrinsic coupling between surface topology and magnetic ordering, effectively delineating the multidimensional landscape that governs optimal material functionality.

**Conclusion**

To summarize, we demonstrated an autonomous scanning probe microscopy workflow for combinatorial materials libraries, using a ternary $Au-Co-Ni$ thin-film gradient as a model system. We integrated automated AFM/MFM imaging with a bag-of-features representation that encodes nanoscale morphology and magnetic response and coupled this representation to a multi-objective Bayesian optimization scheme. The resulting framework treats each measurement as a vector of robust, physically motivated descriptors, builds surrogate models for these feature objectives, and uses them to guide closed-loop exploration of composition space. We showed that this approach can efficiently reconstruct complex structure–property landscapes, identify regions with distinct combinations of morphological and magnetic behavior, and reveal trade-offs among different nanoscale features via the Pareto front.

Although demonstrated here for $Au-Co-Ni$ and AFM/MFM, the underlying methodology is broadly applicable to other material systems, imaging modalities, and property spaces. Any combinatorial platform that yields spatially resolved data including topography, phase, spectroscopic maps, or multimodal images, can be mapped into a bag-of-features space tailored to the relevant functionalities, and explored using the same multi-objective optimization principles. Likewise, the feature set can be extended to include electronic, mechanical, or electrochemical descriptors, and the optimization targets can be adapted to user-defined scientific or technological goals. In this sense, the framework provides a general recipe for turning complex, high-dimensional microscopy outputs into actionable input for autonomous discovery across diverse materials classes.

More broadly, this work illustrates how intelligent probe microscopy can help close the long-standing gap between high-throughput synthesis and actionable understanding in combinatorial materials science. By combining autonomous SPM control, feature-based representations, and multi-objective Bayesian optimization, we move from static, retrospective analysis of large imaging datasets to dynamic, model-driven exploration in which nanoscale measurements directly inform where to look next. As automated experiments, digital twins, and self-driving laboratories continue to develop, we anticipate that similar workflows will become central tools for mapping structure–property relationships, discovering emergent functionalities, and navigating the vast design spaces that define modern materials research.

**ACKNOWLEDGMENTS:**

This research is supported (K.B, H.Z, P.R, and S.V.K) by the National Science Foundation Materials Research Science and Engineering Center program through the UT Knoxville Center for Advanced Materials and Manufacturing (DMR-2309083); (D.A.G and C.C.B acknowledge funding from the US Department of Energy under award DE-SC0026383.

**AUTHOR DECLARATIONS**

Conflict of Interest: The authors have no conflicts to disclose.

## Materials and Methods

### Sample Fabrication and Characterization

A $[Au]_x[Co]_y[Ni]_{(1-x-y)}$ thin film was co-sputtered on 100 mm diameter $SiO_2$ (500 nm) coated $Si$ substrates from pure $Au$, $Co$, and $Ni$ targets to form an $[Au]_x[Co]_y[Ni]_{(1-x-y)}$ compositional gradient, as visualized by the elemental composition map in Figure. 1. The sputtering powers for $Au$, $Co$, and $Ni$ were $4W$ ($DC$), $160W$ ($RF$), and $20W$ ($DC$), respectively, and the deposition time was 120 minutes. Pressure in the sputtering chamber was reduced to below $2 * 10^{-6}$ $Torr$, and sputtering was carried out at an $Ar$ pressure of $5 mTorr$. X-ray diffraction (XRD) patterns were collected using a Malvern Panalytical X'Pert3 MRD diffractometer. All measurements were performed using Cu Kα radiation ($\lambda = 0.154\ nm$) in a $2\theta - \omega$ scan with a 3° offset. Scans were acquired over a $2\theta$ range of 20° - 85° with a step size of 0.05°. Energy dispersive X-ray spectroscopy (EDS) and scanning electron microscopy (SEM) were conducted on a Thermo Scientific Helios 5 Hydra UX plasma focused ion beam (PFIB) system. SEM imaging and EDS analysis were performed at an accelerating voltage of $5.00\ keV$ and a beam current of $0.80\ nA$.

### Implementation of Autonomous Experiment

Bayesian Optimization of the bag-of-features requires both sufficient computational resources and an interface library capable of directly controlling the scanning probe microscope (SPM). For this purpose, a Python-based interface library was utilized to automate operation of a Jupiter SPM system (Oxford Instruments Asylum Research) from a local machine without GPU support. While the current setup is optimized for local execution, the framework is compatible with high-performance computing (HPC) systems. Integration with HPC would allow the implementation of more advanced and computationally demanding optimization strategies. Beyond remote control, the library provides programmatic access to intermediate experimental data, such as trace and retrace scan lines across multiple channels, thereby offering a level of functionality comparable to that of manual human operation.[54]

**References:**

1. R. Potyrailo, K. Rajan, K. Stoewe, I. Takeuchi, B. Chisholm and H. Lam, Combinatorial and high-throughput screening of materials libraries: review of state of the art, ACS combinatorial science 13 (6), 579-633 (2011).
2. I. Takeuchi, H. Chang, C. Gao, P. Schultz, X.-D. Xiang, R. Sharma, M. Downes and T. Venkatesan, Combinatorial synthesis and evaluation of epitaxial ferroelectric device libraries, Applied Physics Letters 73 (7), 894-896 (1998).
3. B. Lee, S. Lee, H. G. Jeong and K.-S. Sohn, Solid-state combinatorial screening of (Sr, Ca, Ba, Mg) 2Si5N8: Eu2+ phosphors, ACS Combinatorial Science 13 (2), 154-158 (2011).
4. S. Senkan, Combinatorial heterogeneous catalysis—a new path in an old field, Angewandte Chemie International Edition 40 (2), 312-329 (2001).
5. T. Gebhardt, D. Music, T. Takahashi and J. M. Schneider, Combinatorial thin film materials science: From alloy discovery and optimization to alloy design, Thin Solid Films 520 (17), 5491-5499 (2012).
6. K. Shahzad, A. I. Mardare and A. W. Hassel, Accelerating materials discovery: combinatorial synthesis, high-throughput characterization, and computational advances, Science and Technology of Advanced Materials: Methods 4 (1), 2292486 (2024).
7. Y. G. Shin and R. B. van Breemen, Analysis and screening of combinatorial libraries using mass spectrometry, Biopharmaceutics & drug disposition 22 (7-8), 353-372 (2001).
8. K. Seok Oh, D. K. Kim, W. F. Maier and S. I. Woo, Discovery of new heterogeneous catalysts for the selective oxidation of propane to acrolein, Combinatorial Chemistry & High Throughput Screening 10 (1), 5-12 (2007).
9. M. Murakami, S. Fujino, S.-H. Lim, L. Salamanca-Riba, M. Wuttig, I. Takeuchi, B. Varughese, H. Sugaya, T. Hasegawa and S. E. Lofland, Microstructure and phase control in Bi–Fe–O multiferroic nanocomposite thin films, Applied physics letters 88 (11) (2006).
10. J. M. Zielinski, J. J. Luke, S. Guglietta and C. Krieg, High throughput multi-omics approaches for clinical trial evaluation and drug discovery, Frontiers in immunology 12, 590742 (2021).
11. A. Ludwig, Discovery of new materials using combinatorial synthesis and high-throughput characterization of thin-film materials libraries combined with computational methods, npj computational materials 5 (1), 70 (2019).
12. Y. Iwasaki, R. Sawada, V. Stanev, M. Ishida, A. Kirihara, Y. Omori, H. Someya, I. Takeuchi, E. Saitoh and S. Yorozu, Identification of advanced spin-driven thermoelectric materials via interpretable machine learning, npj Computational Materials 5 (1), 103 (2019).
13. Z. Wu, S. J. Kan, R. D. Lewis, B. J. Wittmann and F. H. Arnold, Machine learning-assisted directed protein evolution with combinatorial libraries, Proceedings of the National Academy of Sciences 116 (18), 8852-8858 (2019).
14. A. J. Gormley and M. A. Webb, Machine learning in combinatorial polymer chemistry, Nature Reviews Materials 6 (8), 642-644 (2021).
15. M. A. Ziatdinov, Y. Liu, A. N. Morozovska, E. A. Eliseev, X. Zhang, I. Takeuchi and S. V. Kalinin, Hypothesis learning in automated experiment: application to combinatorial materials libraries, Advanced Materials 34 (20), 2201345 (2022).
16. E. Tourani, B. J. Edwards and B. Khomami, Directional entropy bands for surface characterization of polymer crystallization, Polymers 17 (17), 2399 (2025).
17. E. Tourani, B. J. Edwards and B. Khomami, Machine learning workflow for analysis of high-dimensional order parameter space: A case study of polymer crystallization from molecular dynamics simulations, The Journal of Chemical Physics 163 (16) (2025).

18. R. Baptista and M. Poloczek, presented at the International conference on machine learning, 2018 (unpublished).
19. T. Ueno, T. D. Rhone, Z. Hou, T. Mizoguchi and K. Tsuda, COMBO: An efficient Bayesian optimization library for materials science, Materials discovery 4, 18-21 (2016).
20. A. G. Kusne, H. Yu, C. Wu, H. Zhang, J. Hattrick-Simpers, B. DeCost, S. Sarker, C. Oses, C. Toher and S. Curtarolo, On-the-fly closed-loop materials discovery via Bayesian active learning, Nature communications 11 (1), 5966 (2020).
21. Y. Hao, L. Duo and J. He, Autonomous materials synthesis laboratories: Integrating artificial intelligence with advanced robotics for accelerated discovery,  (2025).
22. Y. Liu, R. Pant, I. Takeuchi, R. J. Spurling, J.-P. Maria, M. Ziatdinov and S. V. Kalinin, Automated Materials Discovery Platform Realized: Scanning Probe Microscopy of Combinatorial Libraries, arXiv preprint arXiv:2412.18067 (2024).
23. K. Barakati, Y. Liu, H. Funakubo and S. V. Kalinin, Exploring Domain-Wall Pinning in Ferroelectrics via Automated High-Throughput Atomic Force Microscopy, ACS Applied Materials & Interfaces (2025).
24. R. K. Vasudevan, K. Choudhary, A. Mehta, R. Smith, G. Kusne, F. Tavazza, L. Vlcek, M. Ziatdinov, S. V. Kalinin and J. Hattrick-Simpers, Materials science in the AI age: high-throughput library generation, machine learning and a pathway from correlations to the underpinning physics, MRS communications 9 (3), 10.1557/mrc. 2019.1595 (2019).
25. S. V. Kalinin, M. Ziatdinov, S. R. Spurgeon, C. Ophus, E. A. Stach, T. Susi, J. Agar and J. Randall, Deep learning for electron and scanning probe microscopy: From materials design to atomic fabrication, MRS Bulletin 47 (9), 931-939 (2022).
26. S. V. Kalinin, M. Ziatdinov, J. Hinkle, S. Jesse, A. Ghosh, K. P. Kelley, A. R. Lupini, B. G. Sumpter and R. K. Vasudevan, Automated and autonomous experiments in electron and scanning probe microscopy, ACS nano 15 (8), 12604-12627 (2021).
27. M. Pelikan, in *Hierarchical Bayesian optimization algorithm: toward a new generation of evolutionary algorithms* (Springer, 2005), pp. 31-48.
28. S.-H. Whang, *Nanostructured metals and alloys: processing, microstructure, mechanical properties and applications*. (Elsevier, 2011).
29. M. McHenry and D. Laughlin, Nano-scale materials development for future magnetic applications, Acta materialia 48 (1), 223-238 (2000).
30. J. Staunton, S. Razee, M. F. Ling, D. D. Johnson and F. Pinski, Magnetic alloys, their electronic structure and micromagnetic and microstructural models, Journal of Physics D: Applied Physics 31 (19), 2355 (1998).
31. H. P. Wijn, *Magnetic Properties of Metals: d-elements, alloys and Compounds*. (Springer Science & Business Media, 2012).
32. S. Armyanov, Crystallographic structure and magnetic properties of electrodeposited cobalt and cobalt alloys, Electrochimica Acta 45 (20), 3323-3335 (2000).
33. M. Milyaev, N. Bannikova, L. Naumova, V. Proglyado, E. Patrakov, N. Glazunov and V. Ustinov, Effective Co-rich ternary CoFeNi alloys for spintronics application, Journal of Alloys and Compounds 854, 157171 (2021).
34. M. Shiga, Magnetic properties of Fe65 (Ni1− xMnx) 35 ternary alloys, journal of the physical society of japan 22 (2), 539-546 (1967).
35. L. Betancourt-Cantera, F. Sánchez-De Jesús, A. Bolarín-Miró, A. Gallegos-Melgar, J. Mayen and J. Betancourt-Cantera, Structural analysis and magnetic characterization of ternary alloys (Co–Fe–Ni) synthesized by mechanical alloying, Journal of Materials Research and Technology 9 (6), 14969-14978 (2020).

36. T. McGuire and R. Potter, Anisotropic magnetoresistance in ferromagnetic 3d alloys, IEEE Transactions on magnetics 11 (4), 1018-1038 (2003).
37. C. Chappert and P. Bruno, Magnetic anisotropy in metallic ultrathin films and related experiments on cobalt films, Journal of Applied Physics 64 (10), 5736-5741 (1988).
38. M. Weiler, L. Dreher, C. Heeg, H. Huebl, R. Gross, M. S. Brandt and S. T. Goennenwein, Elastically driven ferromagnetic resonance in nickel thin films, Physical review letters 106 (11), 117601 (2011).
39. R. Collette, Y. Wu and P. D. Rack, Correlating the optical property evolution in the Au-Ni binary thin films: From metastable solid solution to phase separated alloy, Journal of Alloys and Compounds 793, 695-704 (2019).
40. W. L. Boldman, D. A. Garfinkel, R. Collette, C. S. Jorgenson, D. K. Pradhan, D. A. Gilbert and P. D. Rack, Exploring the composition, phase separation and structure of AgFe alloys for magneto-optical applications, Materials Science and Engineering: B 266, 115044 (2021).
41. D. A. Garfinkel, N. Tang, G. Pakeltis, R. Emery, I. N. Ivanov, D. A. Gilbert and P. D. Rack, Magnetic and optical properties of Au–Co solid solution and phase-separated thin films and nanoparticles, ACS Applied Materials & Interfaces 14 (13), 15047-15058 (2022).
42. Y. Wu, C. Liu, T. M. Moore, G. A. Magel, D. A. Garfinkel, J. P. Camden, M. G. Stanford, G. Duscher and P. D. Rack, Exploring photothermal pathways via in situ laser heating in the transmission electron microscope: recrystallization, grain growth, phase separation, and dewetting in Ag0. 5Ni0. 5 Thin Films, Microscopy and microanalysis 24 (6), 647-656 (2018).
43. N. Tang, L. Quigley, W. L. Boldman, C. S. Jorgensen, R. Koch, D. O'Leary, H. R. Medal, P. D. Rack and D. A. Gilbert, Magnetism in metastable and annealed compositionally complex alloys, Physical Review Materials 5 (11), 114405 (2021).
44. V. Chawla, S. Puplampu, H. Zhu, P. D. Rack, D. Penumadu and S. Kalinin, Accelerated Materials Discovery through Cost-Aware Bayesian Optimization of Real-World Indentation Workflows, arXiv preprint arXiv:2511.16930 (2025).
45. M. Ziatdinov, Y. Liu, K. Kelley, R. Vasudevan and S. V. Kalinin, Bayesian active learning for scanning probe microscopy: From Gaussian processes to hypothesis learning, ACS nano 16 (9), 13492-13512 (2022).
46. S. J. Pennycook and C. Colliex, Spectroscopic imaging in electron microscopy, MRS Bulletin 37 (1), 13-18 (2012).
47. S. Jesse and S. V. Kalinin, Principal component and spatial correlation analysis of spectroscopic-imaging data inscanning probe microscopy, Nanotechnology 20 (8), 085714 (2009).
48. P. Batson, Simultaneous STEM imaging and electron energy-loss spectroscopy with atomic-column sensitivity, Nature 366 (6457), 727-728 (1993).
49. U. Pratiush, K. M. Roccapriore, Y. Liu, G. Duscher, M. Ziatdinov and S. V. Kalinin, Building workflows for an interactive human-in-the-loop automated experiment (hAE) in STEM-EELS, Digital Discovery 4 (5), 1323-1338 (2025).
50. R. Liu, H. Pirie, C. Matt, P. Chen, J. E. Hoffman and S. V. Kalinin,  (Oxford University Press US, 2024).
51. U. Pratiush, G. Duscher and S. Kalinin, presented at the AI for Accelerated Materials Design-NeurIPS 2024, 2024 (unpublished).
52. R. Bulanadi, J. Chowdhury, H. Funakubo, M. Ziatdinov, R. Vasudevan, A. Biswas and Y. Liu, Beyond optimization: Exploring novelty discovery in autonomous experiments, ACS Nanoscience Au (2025).
53. Y. Liu, K. P. Kelley, R. K. Vasudevan, H. Funakubo, M. A. Ziatdinov and S. V. Kalinin, Experimental discovery of structure–property relationships in ferroelectric materials via active learning, Nature Machine Intelligence 4 (4), 341-350 (2022).

54. Y. Liu, U. Pratiush, J. Bemis, R. Proksch, R. Emery, P. D. Rack, Y.-C. Liu, J.-C. Yang, S. Udovenko and S. Trolier-McKinstry, Integration of scanning probe microscope with high-performance computing: Fixed-policy and reward-driven workflows implementation, Review of Scientific Instruments 95 (9) (2024).